\documentclass[preprint,12pt, a4paper]{elsarticle}

\usepackage[T1]{fontenc}
\usepackage[utf8]{inputenc}

\usepackage{amssymb}
\usepackage[boxed]{algorithm2e}
\usepackage[numbers]{natbib}
\usepackage{mathtools}
\usepackage[scientific-notation=true]{siunitx}
\usepackage{framed} 
\usepackage{multicol} 
\usepackage{listings}
\usepackage{jlcode}
\usepackage{hyperref}

\hypersetup{
  colorlinks=true
}

\usepackage{lineno}

\usepackage{float}
\restylefloat{table}

\SetKwInput{KwInput}{Input}
\SetKwInput{KwOutput}{Output}

\journal{SoftwareX}

\begin{document}

\begin{frontmatter}



\title{AnyMOD.jl: A Julia package for creating energy system models}


\author{L. Göke}

\address{TU Berlin, Workgroup for Infrastructure Policy (WIP), Straße des 17. Juni 135, 10623 Berlin, Germany.}

\begin{abstract}
AnyMOD.jl is a Julia framework for creating large-scale energy system models with multiple periods of capacity expansion. It applies a novel graph-based approach that was developed to address the challenges in modeling high levels of intermittent generation and sectoral integration. Created models are formulated as linear optimization problems using JuMP.jl as a backend.

To enable modelers to work more efficiently, the framework provides additional features that help to visualize results, streamline the read-in of input data, and rescale optimization problems to increase solver performance.

\end{abstract}


\end{frontmatter}

\section*{Current code version}

\begin{table}[H]
\begin{tabular}{|l|p{6.5cm}|p{6.5cm}|}
\hline
\textbf{Nr.} & \textbf{Code metadata description} & \textbf{Please fill in this column} \\
\hline
C1 & Current code version & v0.1.6 \\
\hline
C2 & Permanent link to code/repository used for this code version & \url{https://github.com/leonardgoeke/AnyMOD.jl/releases/tag/v0.1.6} \\
\hline
C3 & Code Ocean compute capsule & \\
\hline
C4 & Legal Code License   & MIT license (MIT) \\
\hline
C5 & Code versioning system used & git \\
\hline
C6 & Software code languages, tools, and services used & Julia \\
\hline
C7 & Compilation requirements, operating environments \& dependencies & Julia 1.3.1 \\
\hline
C8 & If available Link to developer documentation/manual & \url{https://leonardgoeke.github.io/AnyMOD.jl/stable/} \\
\hline
C9 & Support email for questions &  \url{lqo@wip.tu-berlin.de} \\
\hline
\end{tabular}
\caption{Code metadata}
\end{table}

\linenumbers

\section{Motivation and significance}

Since the production of energy accounts for three-quarters of global emissions, mitigating climate change requires the decarbonization of the energy system \citep{IPCC2014}. Cutting emissions requires to shift supply of primary energy to electricity from wind and solar and extend its use to other sectors. As a result, the energy system has to undergo fundamental change and evolve from largely independent sectors with little supply from renewables into an integrated system characterized by fluctuating renewables.

Capacity expansion models investigate the long-term developments of macro-energy systems, but existing methods were developed for systems still characterized by fossil fuels and struggle to describe the transformation towards a renewable system \citep{Levi2019}. Models like ReEDS, Message, or Switch, pursue a time-slice approach, that reduces the entire year to a small number of independent periods \citep{reeds,Howells2011,switch}. This reduction limits the detail applied to fluctuating renewables and more importantly prohibits to consider long-term storage, a key component of renewable energy systems \citep{Schill2020,Goeke2021}. Other models, like PyPSA or Calliope, diverge from this approach and consider a continuous and hourly time-series instead, which enables a detailed representation of renewables and long-term storage \citep{Pfenninger2018,PyPSA}. But in return these models are limited to a single year and, opposed to models using time-slices, cannot analyze development pathways for today's system.

Against this background, AnyMOD.jl provides a framework for modeling the long-term transformation of the energy system with the level of detail necessary to represent fluctuating renewables and long-term storage. The framework implements a novel graph-based method introduced in \citet{Goeke2020} that varies the level of temporal and spatial detail by energy carrier to keep models with high resolution computationally tractable. The approach also enables to model the substitution of energy carriers and, on the practical side, facilitates the read-in of input data.

AnyMOD.jl follows an easy to use, but difficult to master principle. Since individual models are solely defined by CSV files and can be run with a few lines of standard code, running an existing model, and performing sensitivity analysis requires little experience. More advanced applications, like creating new models and individually modifying their formulation, requires some programming skills and a deeper understanding of the framework's structure. Since models are defined from CSV files and short code scripts, the framework supports version-controlled model development to promote collaboration and transparency.

The following section gives an overview of the framework's structure and presents two functionalities with greater detail, the read-in of parameter data (section \ref{head:1}) and the re-scaling algorithm (section \ref{head:2}). The subsequent section describes an application that models the transformation of the European power and gas sector. The final section paper highlights the framework's impact and concludes. 

\section{Software description}

The package is implemented in Julia. Its key dependencies are JuMP.jl as a backend for linear optimization and DataFrames.jl for data processing \citep{Dunning2017, Bezanson2017}. The framework uses PyCall.jl to create an internal Python environment and apply the Python packages NetworkX and Plotly for plotting. Gurobi is added as an optional dependency, because its function to compute irreducible inconsistent subsystems is utilized to debug infeasible models. Apart from that, the framework is compatible with any open or commercial solver implemented in Julia. To increase performance the package heavily utilizes Julia's multi-threading capabilities. Since not supported by JuMP.jl, the mere creation of constraints uses only one thread, but the computationally more intensive composition of constraints from variables and parameters is multi-threaded.

\subsection{Software Architecture}
 
\begin{figure}
	\centering
		\includegraphics[scale=0.28]{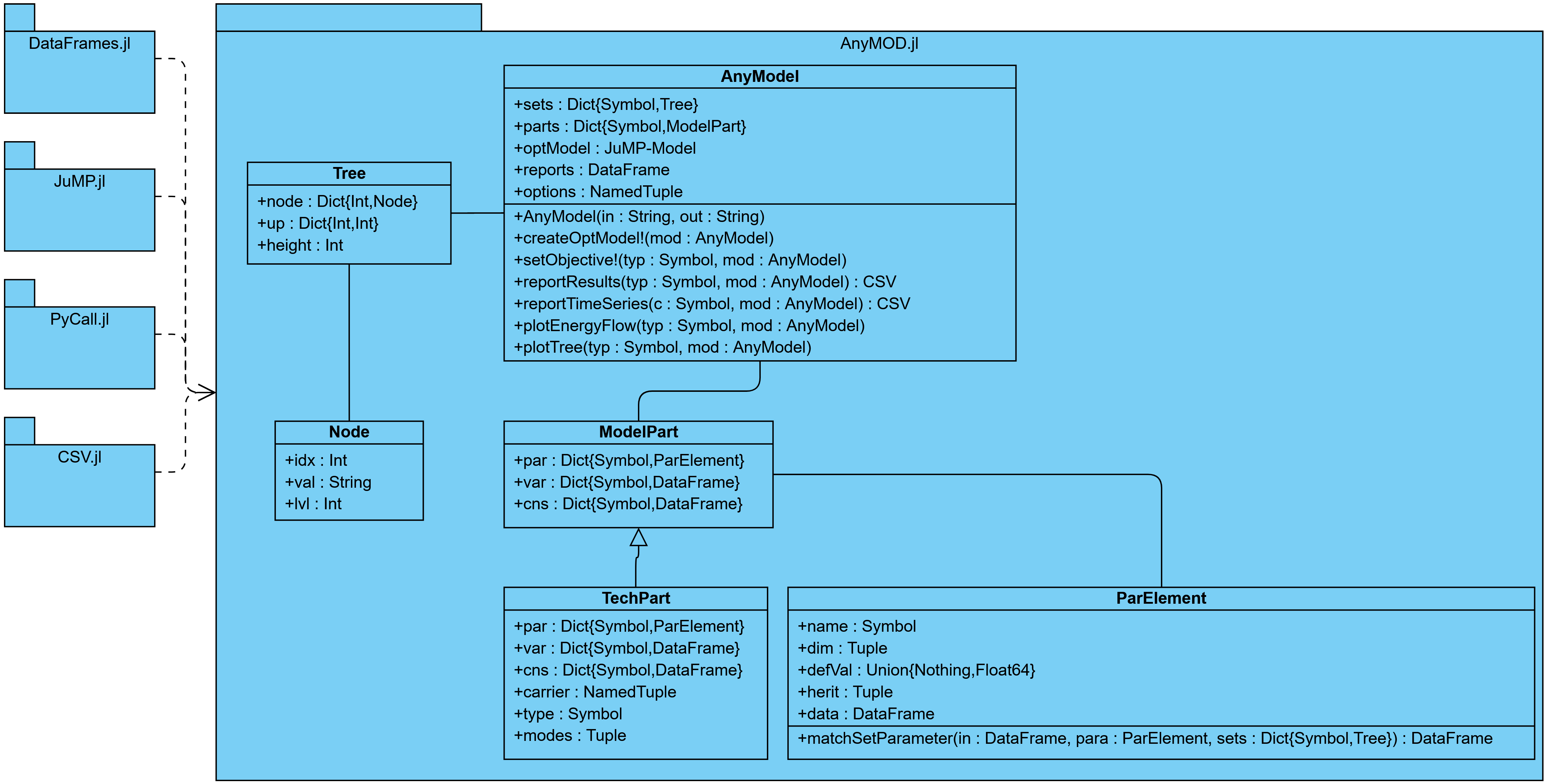}
	\caption{UML class diagram of package components}
	\label{fig:1}
\end{figure}

The class diagram in Figure \ref{fig:1} illustrates the architecture of AnyMOD.jl and how it revolves around the \textit{AnyModel} object. For the sake of clarity, the diagram is not exhaustive and only covers the most relevant dependencies, objects and attributes. Listing \ref{lis:1} provides the corresponding code to initialize, populate, solve and analyze the model object.

After loading AnyMOD.jl, the constructor initializes the \textit{AnyModel} object based on two mandatory arguments: an input directory and an output directory. The CSV files defining a model consist of set and parameter files that have to be placed in the input directory. The set files define all time-steps, regions, energy carriers and technologies considered in a model and map how these are related, for example which carriers a technology can generate. Following the graph-based approach, the elements of each set are organized as nodes of hierarchical trees. 

\begin{lstlisting}[caption={Script to initialize, create and run a model},captionpos=b] 
using AnyMOD # loading packages
model_object = anyModel("../demo","results") # construct model object

# create optimization problem and set an objective
createOptModel!(model_object)
setObjective!(:costs, model_object)

# solve model and report results
using Cbc
set_optimizer(model_object.optModel, Cbc.Optimizer)
optimize!(model_object.optModel)
reportResults(:summary, model_object)
\end{lstlisting}
\label{lis:1}

Qualitative inputs on sets are complemented with quantitative data from the parameter files, that for instance provide demand time-series or technology properties like investment costs or efficiency. While the naming and format of set files is strictly defined, parameter data can be freely structured and distributed across files. As a result, models can be composed modularly, since different models can share the same input files. After reading in all parameter data, the constructor creates a \textit{ParElement} object for each parameter with data and meta information and assigns it to a \textit{ModelPart} object. The \textit{ModelPart} objects partition the model into different parts, for instance, the \textit{ParElement} for demand time-series will be assigned to a model part dedicated to the energy balance. Each technology got its own part object of the subclass \textit{TechPart}, that also stores technology specific attributes like assigned carriers. 
 
After construction, the \textit{AnyModel} object is passed to the \textit{createModel!} function, which creates all the variables and constraints of the underlying optimization problem \textit{optModel}. These variables and constraints are again assigned to model parts and stored as data frames. For instance, Table \ref{tab:1} depicts a data frame of generation variables. The column on the right stores the JuMP variable objects and the four other columns give the time-step, region, carrier, and technology of each variable, which are provided as indexes of the \textit{Node} objects created during initialization. Such data frames for variables are combined with parameter data using database operations to construct constraints. For instance, generation variables are aggregated by technology and than joined with the demand parameter to create the energy balance in Table \ref{tab:2}.

\begin{table}
  \centering
\begin{tabular}{cccc|c}
time-step & region & carrier & technology & variable          \\ \hline
1        & 1      & 1       & 1          & $gen(1,1,1,1)$ \\
2        & 1      & 1       & 1          & $gen(2,1,1,1)$ \\
3        & 1      & 1       & 1          & $gen(3,1,1,1)$
\end{tabular}
\caption{Exemplary data frame of generation variables}%
\label{tab:1}%
\end{table}

\begin{table}
  \centering
\begin{tabular}{ccc|c}
time-step & region & carrier & constraint      \\ \hline
1        & 1      & 1       & $dem(1,1,1) = \sum_{t} gen(1,1,1,t)$ \\
2        & 1      & 1       & $dem(2,1,1) = \sum_{t} gen(2,1,1,t)$    \\
3        & 1      & 1       & $dem(3,1,1) = \sum_{t} gen(3,1,1,t)$                                             
\end{tabular}
\caption{Exemplary data frame of energy balance constraints}%
\label{tab:2}%
\end{table}

After the optimization problem is created, its objective is set with the \textit{setObjective} function. At this point the user can also freely modify and extend the automatically generated problem by accessing the JuMP attributes of the \textit{AnyModel} object and its parts. Finally, the optimization problem \textit{optModel} is passed to a solver and analyzed afterwards. All results are written to the output directory, which was passed to the constructor in the beginning. A reporting file with error messages and warnings is written to this directory as well.

\subsection{Software Functionalities}

As outlined above, AnyMOD.jl is a package for the creation of energy system models. Additional features are aimed either at simplifying its application or enhancing the performance of creating and solving models. In the following, two of these features are presented in greater detail.

\subsubsection{Inheritance Algorithm}
\label{head:1}

As explained above, model constraints are constructed from variables and parameters, which are again defined by input data. Usually, models use a single parameter value in many constraints. For example, the efficiency of a newly build gas power plant does typically not vary by time-step or region and all constraints describing these plants will use the same value. Consequently, it would be inefficient, if AnyMOD.jl required users to provide efficiency data at a temporal and spatial resolution. On the other hand, efficiencies of heat-pumps are highly dependant on region and time-step, because they depend on ambient temperature. So, not permitting efficiencies to depend on time-step and region, would prevent to model these technologies accurately. A similar problem occurs, if investment costs of emerging technologies, like photovoltaic, are expected to decrease within the model horizon, but costs for other technologies remain constant. Here, providing all costs at a yearly resolution leads to redundant inputs for most technologies, but if costs cannot be varied by year at all, cost degression of photovoltaic cannot be modelled. In conclusion a predefined resolution of input data either results in an highly inefficient read-in of input data or restricts modelling capabilities. 

To resolve this problem, AnyMOD.jl does not predefine the resolution of input data and instead automatically infers how data should be used from the way it is specified. For example, providing different efficiencies in dependence of time-step and region will result in temporally and spatially resolved efficiencies in the model, but if instead a parameter is provided without time-steps or regions, the model uses a uniform value. This concept is not limited to certain parameters or dimensions, but applies comprehensively. The implementing algorithm builds on the idea to "inherit" missing data for a specific node from its relatives in the hierarchical tree.\footnote{This idea of "inheritance" is not be confused with inheritance in the context of object orientated programming.}

\begin{figure}
	\centering
		\includegraphics[scale=0.25]{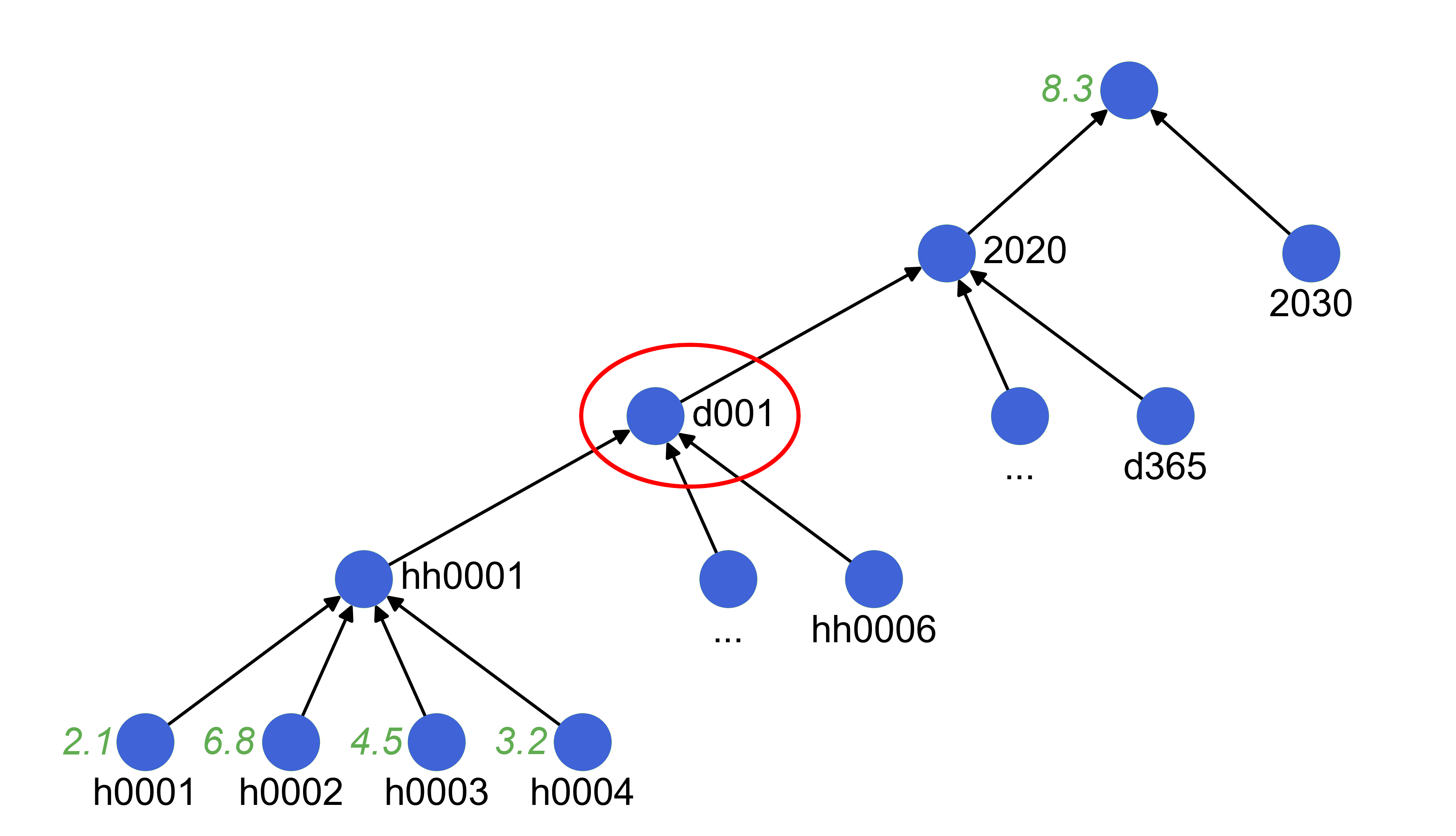}
	\caption{Basic mechanism of inheritance within hierarchical trees}
	\label{fig:2}
\end{figure}

Figure \ref{fig:2} illustrates the basic mechanism of the algorithm based on an exemplary hierarchical tree organizing time-steps. The first level of the tree organizes different years with days, 4-hours steps and hours following on the subsequent levels. Green numbers indicate input data provided for a specific node. If input data is not specified in dependence of the time-step, it is assigned to the root of the tree. The algorithm can obtain missing data at the circled node in three different ways: either move up the tree and use '8.3', move down the tree and sum the hourly values, or move down the tree and average the hourly values. 
How the algorithm deploys these three methods for each dimension depends on the inheritance rules of the parameter. A detailed overview for each parameter is provided in the \href{https://leonardgoeke.github.io/AnyMOD.jl/stable/parameter_list}{parameter list} of the documentation.

Figure \ref{fig:3} outlines how these rules then are deployed to obtain parameter data. The described algorithm corresponds to the \textit{matchSetParameter} function of \textit{ParElement} in Figure \ref{fig:1} and takes the following inputs: a data frame to be filled with parameter data, a respective parameter object and the hierarchical trees. First, the algorithm checks for direct matches between the input data frame and the parameter data. Afterwards, it loops over the inheritance rules to inherit new data for missing nodes as described above. If new data is obtained, the algorithm checks again for matches with the input data frame. The loops ends when either all rows are matched with data, or all inheritance rules have been applied. In the latter case, unassigned rows are either dropped or assigned a default value, if one is defined for the respective parameter.

\begin{figure}[h]
\begin{algorithm}[H]
\KwInput{data frame requiring data, parameter object, hierarchical trees}
\KwOutput{data frame with parameters assigned}
    find matches of data frame with parameter data\;
    \For{I}{
        try to inherit new data for missing nodes\;
        \If{new data obtained}{
        add newly obtained data to parameter object\;
        find new matches of data frame and parameter data\;
                \If{no unmatched rows in data frame anymore}{
        exit loop\;}
        }

    }
    \texttt{\\}
    \uIf{parameter has default value}{
    use default for unmatched rows\;}
    \Else{
    drop unmatched rows\;}
\end{algorithm}
\caption{Inheritance algorithm}
\label{fig:3}
\end{figure}

\subsubsection{Scaling}
\label{head:2}

The formulation of an optimization problem can have a major impact on solver performance. The barrier algorithm, the fastest method for solving large linear problems, is particularly sensitive to a model's numerical properties, and poor formulations will thus greatly increase computation time. For this reason, AnyMOD.jl automatically applies a two-step scaling process when creating optimization problems. The process aims to narrow the range of coefficients and constants in a problem between $10^{-3}$ and $10^{6}$, as recommended.\footnote{See the \href{http://www.gurobi.com/documentation/9.0/refman/num_grb_guidelines_for_num.html}{Gurobi Guidelines for Numerical Issues} for details.}

As a demonstration of how this range is achieved, Eq. \ref{eq:1} constitutes the constraints of an exemplary linear model. In the first and second row, the coefficients for $x_1$ are currently outside of the targeted interval. In addition, the maximum range of coefficients in the second row amounts to $10^{11}$ (= $\frac{10^{2}}{10^{-9}}$), which exceeds the maximum range of the targeted interval of $10^{9}$ (= $\frac{10^{6}}{10^{-3}}$) and means the equation cannot be multiplied with a constant factor to shift coefficients into the desired interval.

\begin{equation}  \label{eq:1}
\begin{array}[t]{clcclcclcclcc}
10^{-8} & x_1  & + &  10^{3} & x_2 & + & & x_3 & \leq & & b_1 &\\
10^{-9} & x_1  & + &  10^{2}& x_2 & + & &  x_3 & \leq & &  b_2 & \\ 
& x_1  & + &  &  x_2 & + &  &  x_3 & \leq & & b_2 &
\end{array}
\end{equation}

Therefore, in the first step the maximum range of coefficients is decreased by substituting variables. In the example, $x_1$ is substituted with $10^3 \, x'_1$, which results in the system displayed in Eq. \ref{eq:2}. 
\begin{equation} \label{eq:2}
\begin{array}[t]{lcclcclcclcc}
10^{-5} & x'_1  & + &  10^{3} & x_2 & + & & x_3 & \leq & & b_1 &\\
10^{-6} & x'_1  & + &  10^{2}& x_2 & + & &  x_3 & \leq & & b_2 &\\
10^{3} & x'_1  & + &  &  x_2 & + &  &  x_3 & \leq & & b_2 &
\end{array}
\end{equation}

Since the first step decreased the maximum range, in the second step coefficients can be shifted into the interval between $10^{-3}$ and $10^{6}$. For this purpose, each constraint (or row) is scaled with a constant factor. In the example, the first row is multiplied by $10^2$ and the second row by $10^3$ resulting in the system displayed in Eq. \ref{eq:3} that finally complies with the recommended range.   
\begin{equation} \label{eq:3}
\begin{array}[t]{lcclcclcclcc}
10^{-3} & x'_1  & + &  10^{5} & x_2 & + & 10^{2} & x_3 & \leq & 10^{2} & b_1 & \\
10^{-3} & x'_1  & + &  10^{5}& x_2 & + & 10^{3} &  x_3 & \leq & 10^{3} & b_2 & \\
10^{3} & x'_1  & + &  &  x_2 & + &  &  x_3 & \leq & & b_2 &\\
\end{array}
\end{equation}
AnyMOD.jl uses default factors for substitution that depend on the variable type and can be adjusted if they fail to achieve the desired result. Factors for scaling can be automatically computed based on the current coefficients in a constraint.

\begin{figure} 
	\centering
		\includegraphics[scale=0.53]{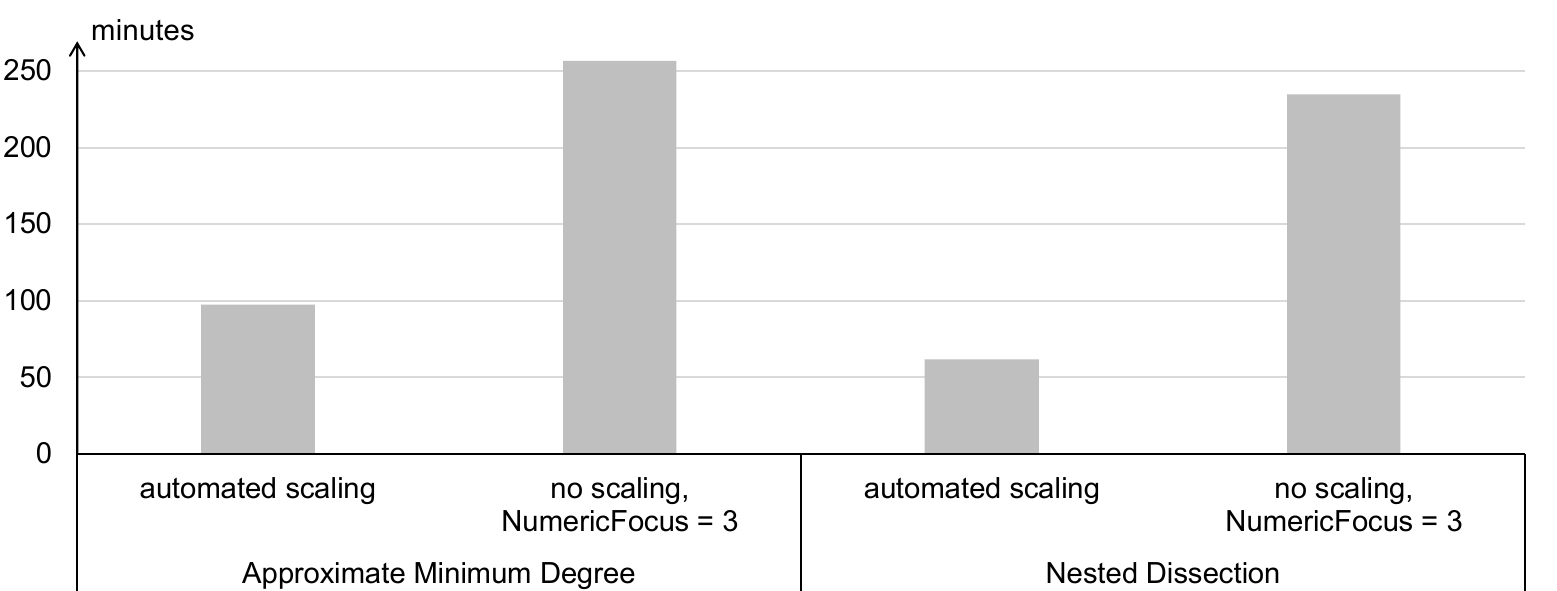}
	\caption{Impact of scaling algorithm on solver run-time}
\label{fig:4}
\end{figure}

Figure \ref{fig:4} demonstrates the impact of automated scaling by comparing the solve times of Gurobi's barrier implementation for a test model.\footnote{The corresponding model files can be found in the following repository: \url{https://github.com/leonardgoeke/AnyMOD_example_model/tree/May2020}} To ensure robustness of the results, Barrier was run with both available ordering algorithms, “approximate minimum degree” and “nested dissection.” With automated scaling disabled, a \textit{NumericFocus} parameter of three is necessary to avoid early termination or extremely long solve times due to numerical difficulties. In conclusion, automated scaling decreases solve time of the test model roughly by a factor of three.

\section{Illustrative Example}
\label{head:3}

\citet{Hainsch2020b} applied AnyMOD.jl to the decarbonization of the European power and gas sector on a pathway from 2030 to 2040 instead of a single year. The analysis with AnyMOD.jl complements results from another energy system model with less spatiotemporal detail. The application subdivides Europe on a country level and includes an aggregated representation of transmission infrastructure to enable the exchange of energy carriers between countries.

\begin{figure}
	\centering
		\includegraphics[scale=0.55]{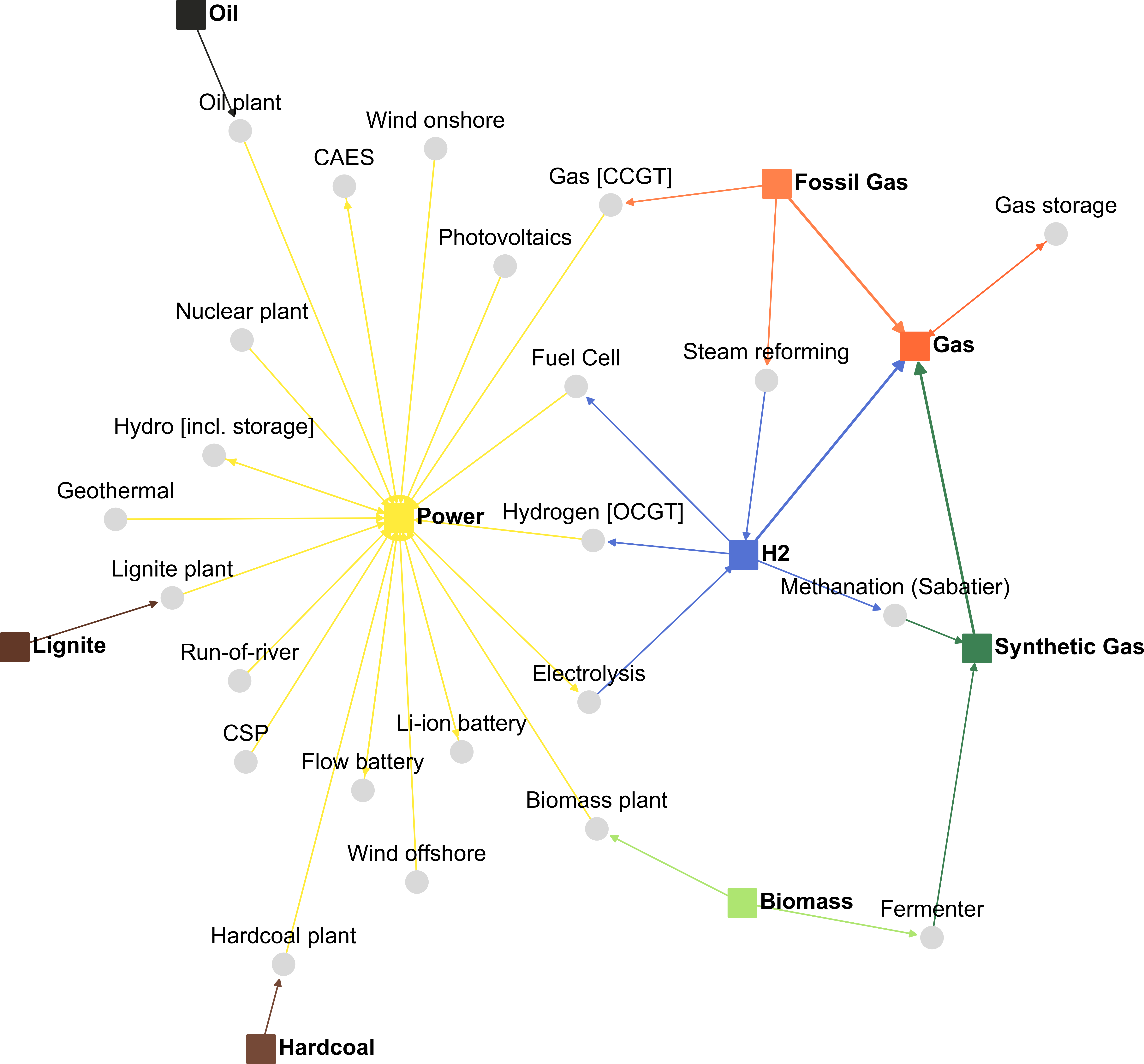}
	\caption{Graph of technologies and energy in example}
	\label{fig:5}
\end{figure}

Figure \ref{fig:5} was plotted with the \textit{plotEnergyFlow} function and provides an overview of the technologies and energy carriers considered. In the graph, carriers are symbolized by colored vertices and technologies by gray vertices. Entering edges of technologies point towards their input carriers; outgoing edges refer to outputs. Since the model includes both the power and gas sector, it is not limited to short-term storage of power, like batteries, but also considers creation and utilization of synthetic fuels for long-term storage. Since fluctuating renewables are the main source of supply by 2040, power is modelled at an hourly resolution. To reduce model size and account for the inherent flexibility of gaseous energy carriers, fossil gas, hydrogen, and synthetic gas are balanced daily instead. All other energy carriers are modelled yearly.

\begin{figure}
	\centering
		\includegraphics[scale=0.28]{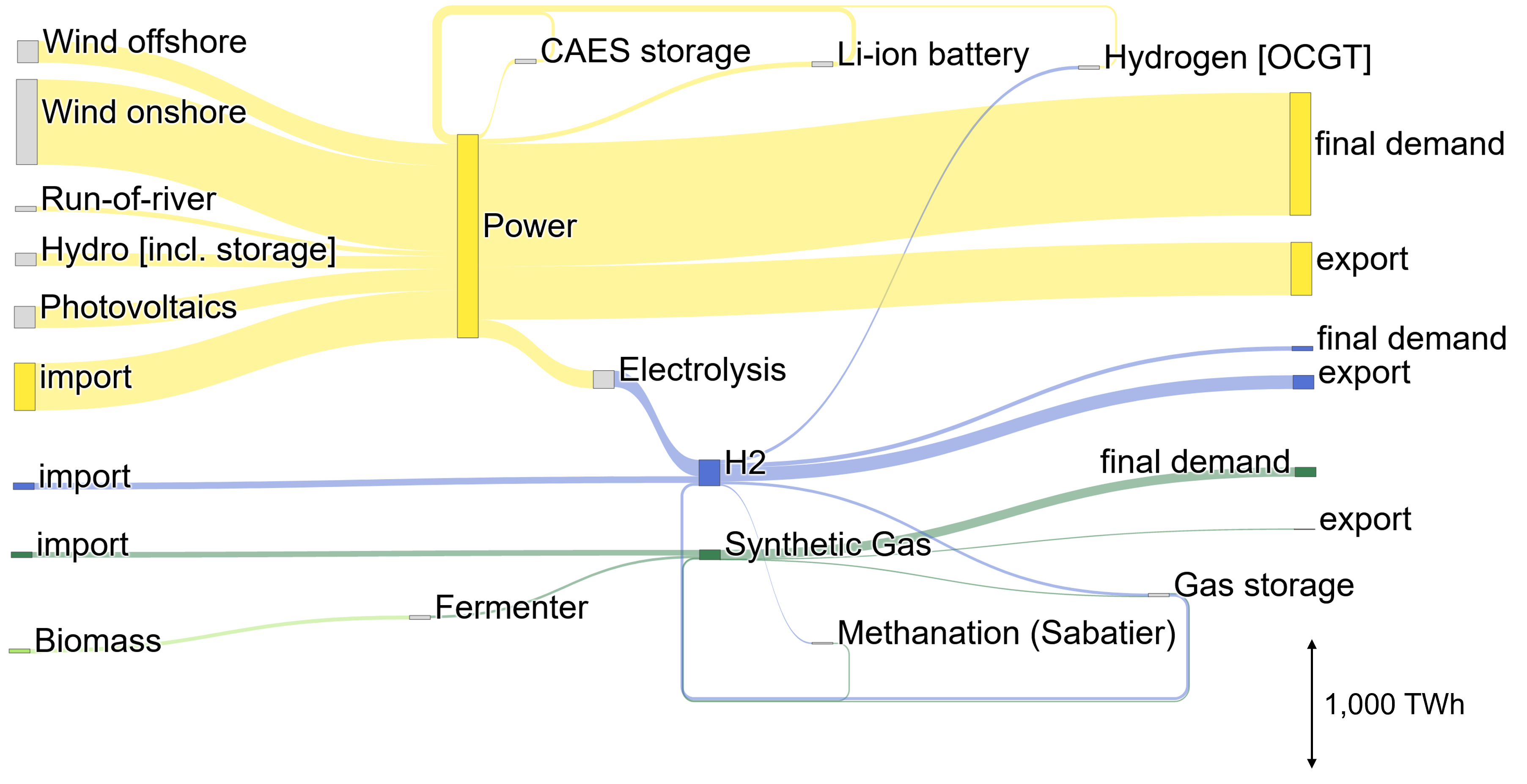}
	\caption{Sankey diagram for France in 2040 in example}
	\label{fig:6}
\end{figure}

The energy flows for France in 2040 when solving the model are displayed in Figure \ref{fig:6}. The sankey diagram does not only show how hydrogen is used for long-term storage of power, but also how final demand for hydrogen and synthetic gases, for example from the industry sector, is covered. In addition, the substantial amount of imports and exports for all carriers highlights the importance of large models that can account for several regions at once.

\section{Impact and conclusions}

AnyMOD.jl provides a framework for modeling the transformation towards a decarbonized energy system at a high spatiotemporal resolution. For this purpose, it implements a graph-based method introduced that enables to vary the level of detail by energy carrier. In addition, the framework introduces a more flexible method to read-in input data and automatically scales created optimization problems to increase solver performance. Lastly, the tool provides advanced plotting features, like Sankey diagrams. 

To facilitate access for users, AnyMOD.jl can be used without any proprietary software. Using the framework does not require extensive programming skills but supports version-controlled model development, since models are created from CSV files. To extend and modify a created model, advanced users can easily access and manipulate its underlying JuMP objects. The organization of input files is highly flexible and eases the creation of new models from existing files. 

In conclusion, AnyMOD.jl enables research to spend less time on coding and data management and more time focusing on the scientific part of their work. Its high level of accessibility also makes AnyMOD.jl suitable for use by companies, regulators, or non-governmental organizations. Finally, AnyMOD.jl promotes openness and transparency in various ways. Due to the relevance of these qualities for public policy, this is of particular importance with energy system models \citep{Pfenninger2017b}.

Additional features currently developed include a more detailed representation of transmission infrastructure and the inclusion of more than one weather year in a single model. The later also includes the development of a distributed solution algorithm to keep the resulting increase in model size manageable. 

\section{Conflict of Interest}

We wish to confirm that there are no known conflicts of interest associated with this publication and there has been no significant financial support for this work that could have influenced its outcome.

\section*{Acknowledgements}

The research leading to these results has received funding from the European Union’s Horizon 2020 research and innovation programme under grant agreement No 773406. Also, I want to thank Mario
Kendziorski and Richard Weinhold for their constructive feedback and introduction to the Julia language. A special thanks goes to all Julia developers.





\bibliographystyle{elsarticle-num-names}
\bibliography{cas-refs}

\begin{thebibliography}{14}
\expandafter\ifx\csname natexlab\endcsname\relax\def\natexlab#1{#1}\fi
\providecommand{\url}[1]{\texttt{#1}}
\providecommand{\href}[2]{#2}
\providecommand{\path}[1]{#1}
\providecommand{\DOIprefix}{doi:}
\providecommand{\ArXivprefix}{arXiv:}
\providecommand{\URLprefix}{URL: }
\providecommand{\Pubmedprefix}{pmid:}
\providecommand{\doi}[1]{\href{http://dx.doi.org/#1}{\path{#1}}}
\providecommand{\Pubmed}[1]{\href{pmid:#1}{\path{#1}}}
\providecommand{\bibinfo}[2]{#2}
\ifx\xfnm\relax \def\xfnm[#1]{\unskip,\space#1}\fi
\bibitem[{Edenhofer et~al.(2014)Edenhofer, Pichs-Madruga, Sokona, Farahani,
  Kadner, Seyboth, Adler, Baum, Brunner, Eickemeier, Kriemann, Savolainen,
  Schlömer, von Stechow, Zwickel, and Minx}]{IPCC2014}
\bibinfo{author}{O.~Edenhofer}, \bibinfo{author}{R.~Pichs-Madruga},
  \bibinfo{author}{Y.~Sokona}, \bibinfo{author}{E.~Farahani},
  \bibinfo{author}{S.~Kadner}, \bibinfo{author}{K.~Seyboth},
  \bibinfo{author}{A.~Adler}, \bibinfo{author}{I.~Baum},
  \bibinfo{author}{S.~Brunner}, \bibinfo{author}{P.~Eickemeier},
  \bibinfo{author}{B.~Kriemann}, \bibinfo{author}{J.~Savolainen},
  \bibinfo{author}{S.~Schlömer}, \bibinfo{author}{C.~von Stechow},
  \bibinfo{author}{T.~Zwickel}, \bibinfo{author}{J.~Minx},
  \bibinfo{title}{Climate Change 2014: Mitigation of Climate Change.
  Contribution of Working Group III to the Fifth Assessment Report of the
  Intergovernmental Panel on Climate Change}, \bibinfo{publisher}{Cambridge
  University Press}, \bibinfo{address}{Cambridge, United Kingdom and New York,
  NY, USA}, \bibinfo{year}{2014}.
\bibitem[{Levi et~al.(2019)Levi, Kurland, Carbajales-Dale, Weyant, Brandt, and
  Benson}]{Levi2019}
\bibinfo{author}{P.~Levi}, \bibinfo{author}{S.~Kurland},
  \bibinfo{author}{M.~Carbajales-Dale}, \bibinfo{author}{J.~Weyant},
  \bibinfo{author}{A.-R. Brandt}, \bibinfo{author}{S.~Benson},
\newblock \bibinfo{title}{Macro-energy systems: Toward a new discipline},
\newblock \bibinfo{journal}{Joule} \bibinfo{volume}{3} (\bibinfo{year}{2019})
  \bibinfo{pages}{2282--2286}.
  \DOIprefix\doi{https://doi.org/10.1016/j.joule.2019.07.017}.
\bibitem[{Cohen et~al.(2019)Cohen, Becker, Bielen, Brown, Cole, Eurek, Frazier,
  Frew, Gagnon, Ho, Jadun, Mai, Mowers, Murphy, Reimers, Richards, Ryan,
  Spyrou, Steinberg, Sun, Vincent, and Zwerling}]{reeds}
\bibinfo{author}{S.~M. Cohen}, \bibinfo{author}{J.~Becker},
  \bibinfo{author}{D.~A. Bielen}, \bibinfo{author}{M.~Brown},
  \bibinfo{author}{W.~J. Cole}, \bibinfo{author}{K.~P. Eurek},
  \bibinfo{author}{A.~Frazier}, \bibinfo{author}{B.~A. Frew},
  \bibinfo{author}{P.~J. Gagnon}, \bibinfo{author}{J.~L. Ho},
  \bibinfo{author}{P.~Jadun}, \bibinfo{author}{T.~T. Mai},
  \bibinfo{author}{M.~Mowers}, \bibinfo{author}{C.~Murphy},
  \bibinfo{author}{A.~Reimers}, \bibinfo{author}{J.~Richards},
  \bibinfo{author}{N.~Ryan}, \bibinfo{author}{E.~Spyrou},
  \bibinfo{author}{D.~C. Steinberg}, \bibinfo{author}{Y.~Sun},
  \bibinfo{author}{N.~M. Vincent}, \bibinfo{author}{M.~Zwerling},
\newblock \bibinfo{title}{Regional energy deployment system (reeds) model
  documentation: Version 2018}  (\bibinfo{year}{2019}). \URLprefix
  \url{https://www.osti.gov/biblio/1505935}. \DOIprefix\doi{10.2172/1505935}.
\bibitem[{Howells et~al.(2011)Howells, Rogner, Strachan, Heaps, Huntington,
  Kypreos, Hughes, Silveira, DeCarolis, Bazillian, and Roehrl}]{Howells2011}
\bibinfo{author}{M.~Howells}, \bibinfo{author}{H.~Rogner},
  \bibinfo{author}{N.~Strachan}, \bibinfo{author}{C.~Heaps},
  \bibinfo{author}{H.~Huntington}, \bibinfo{author}{S.~Kypreos},
  \bibinfo{author}{A.~Hughes}, \bibinfo{author}{S.~Silveira},
  \bibinfo{author}{J.~DeCarolis}, \bibinfo{author}{M.~Bazillian},
  \bibinfo{author}{A.~Roehrl},
\newblock \bibinfo{title}{Osemosys: The open source energy modeling system: An
  introduction to its ethos, structure and development},
\newblock \bibinfo{journal}{Energy Policy} \bibinfo{volume}{39(10)}
  (\bibinfo{year}{2011}) \bibinfo{pages}{5850--5870}.
  \DOIprefix\doi{10.1016/j.enpol.2011.06.033}.
\bibitem[{Johnston et~al.(2019)Johnston, Henriquez-Auba, Maluenda, and
  Fripp}]{switch}
\bibinfo{author}{J.~Johnston}, \bibinfo{author}{R.~Henriquez-Auba},
  \bibinfo{author}{B.~Maluenda}, \bibinfo{author}{M.~Fripp},
\newblock \bibinfo{title}{Switch 2.0: A modern platform for planning
  high-renewable power systems},
\newblock \bibinfo{journal}{Software X} \bibinfo{volume}{10}
  (\bibinfo{year}{2019}) \bibinfo{pages}{100251}.
  \DOIprefix\doi{10.1016/j.softx.2019.100251}.
\bibitem[{Schill(2020)}]{Schill2020}
\bibinfo{author}{W.~Schill},
\newblock \bibinfo{title}{Electricity storage and the renewable energy
  transition},
\newblock \bibinfo{journal}{Joule} \bibinfo{volume}{4} (\bibinfo{year}{2020})
  \bibinfo{pages}{2047--2064}.
  \DOIprefix\doi{https://doi.org/10.1016/j.joule.2020.07.022}.
\bibitem[{Göke and Kendziorski(2021)}]{Goeke2021}
\bibinfo{author}{L.~Göke}, \bibinfo{author}{M.~Kendziorski},
\newblock \bibinfo{title}{The adequacy of time-series reduction for renewable
  energy systems},
\newblock \bibinfo{journal}{Working Paper}  (\bibinfo{year}{2021}). \URLprefix
  \url{https://arxiv.org/abs/2101.06221}.
\bibitem[{Pfenniger and Pickering(2018)}]{Pfenninger2018}
\bibinfo{author}{S.~Pfenniger}, \bibinfo{author}{B.~Pickering},
\newblock \bibinfo{title}{Calliope: a multi-scale energy systems modelling},
\newblock \bibinfo{journal}{Journal of Open Source Software}
  \bibinfo{volume}{3(29)} (\bibinfo{year}{2018}) \bibinfo{pages}{825}.
  \DOIprefix\doi{10.21105/joss.00825}.
\bibitem[{Brown et~al.(2018)Brown, H\"orsch, and Schlachtberger}]{PyPSA}
\bibinfo{author}{T.~Brown}, \bibinfo{author}{J.~H\"orsch},
  \bibinfo{author}{D.~Schlachtberger},
\newblock \bibinfo{title}{{PyPSA: Python for Power System Analysis}},
\newblock \bibinfo{journal}{Journal of Open Research Software}
  \bibinfo{volume}{6} (\bibinfo{year}{2018}). \URLprefix
  \url{https://doi.org/10.5334/jors.188}. \DOIprefix\doi{10.5334/jors.188}.
  \href{http://arxiv.org/abs/1707.09913}{{\tt arXiv:1707.09913}}.
\bibitem[{Göke(2021)}]{Goeke2020}
\bibinfo{author}{L.~Göke},
\newblock \bibinfo{title}{A graph-based formulation for modeling macro-energy
  systems},
\newblock \bibinfo{journal}{Applied Energy} \bibinfo{volume}{301}
  (\bibinfo{year}{2021}) \bibinfo{pages}{117377}.
  \DOIprefix\doi{10.1016/j.apenergy.2021.117377}.
\bibitem[{Dunning et~al.(2017)Dunning, Huchette, and Lubin}]{Dunning2017}
\bibinfo{author}{I.~Dunning}, \bibinfo{author}{J.~Huchette},
  \bibinfo{author}{M.~Lubin},
\newblock \bibinfo{title}{Jump: A modeling language for mathematical
  optimization},
\newblock \bibinfo{journal}{SIAM Review} \bibinfo{volume}{59(2)}
  (\bibinfo{year}{2017}) \bibinfo{pages}{295--320}.
  \DOIprefix\doi{https://doi.org/10.1137/15M1020575}.
\bibitem[{Bezanson et~al.(2017)Bezanson, Edelman, Karpinski, and
  Shah}]{Bezanson2017}
\bibinfo{author}{J.~Bezanson}, \bibinfo{author}{A.~Edelman},
  \bibinfo{author}{S.~Karpinski}, \bibinfo{author}{V.~Shah},
\newblock \bibinfo{title}{Julia: A fresh approach to numerical computing},
\newblock \bibinfo{journal}{SIAM Review} \bibinfo{volume}{59(1)}
  (\bibinfo{year}{2017}) \bibinfo{pages}{65--98}.
  \DOIprefix\doi{https://doi.org/10.1137/141000671}.
\bibitem[{Hainsch et~al.(2020)Hainsch, Göke, Kemfert, Oei, and
  Hirschhausen}]{Hainsch2020b}
\bibinfo{author}{K.~Hainsch}, \bibinfo{author}{L.~Göke},
  \bibinfo{author}{C.~Kemfert}, \bibinfo{author}{P.-Y. Oei},
  \bibinfo{author}{C.~Hirschhausen},
\newblock \bibinfo{title}{European green deal: Using ambitious climate targets
  and renewable energy to climb out of the economic crisis},
\newblock \bibinfo{journal}{DIW Weekly Report} \bibinfo{volume}{28+29}
  (\bibinfo{year}{2020}).
  \DOIprefix\doi{https://doi.org/10.18723/diw_dwr:2020-28-1}.
\bibitem[{Pfenniger et~al.(2017)Pfenniger, DeCarolis, Hirth, Quoilin, and
  Staffel}]{Pfenninger2017b}
\bibinfo{author}{S.~Pfenniger}, \bibinfo{author}{J.~DeCarolis},
  \bibinfo{author}{L.~Hirth}, \bibinfo{author}{S.~Quoilin},
  \bibinfo{author}{I.~Staffel},
\newblock \bibinfo{title}{The importance of open data and software: Is energy
  research lagging behind?},
\newblock \bibinfo{journal}{Energy Policy} \bibinfo{volume}{101}
  (\bibinfo{year}{2017}) \bibinfo{pages}{211--215}.
  \DOIprefix\doi{10.1016/j.enpol.2016.11.046}.

\end{thebibliography}

\section*{Current executable software version}

\begin{table}[!h]
\begin{tabular}{|l|p{6.5cm}|p{6.5cm}|}
\hline
\textbf{Nr.} & \textbf{(Executable) software metadata description} & \textbf{Please fill in this column} \\
\hline
S1 & Current software version & v0.1.6 \\
\hline
S2 & Permanent link to executables of this version  & \url{https://github.com/leonardgoeke/AnyMOD.jl/releases/tag/v0.1.6} \\
\hline
S3 & Legal Software License & MIT license (MIT) \\
\hline
S4 & Computing platforms/Operating Systems & Linux, Microsoft Windows, iOS \\
\hline
S5 & Installation requirements \& dependencies & Julia 1.3.1\\
\hline
S6 & If available, link to user manual - if formally published include a reference to the publication in the reference list &  \url{https://leonardgoeke.github.io/AnyMOD.jl/stable/} \\
\hline
S7 & Support email for questions & \url{lqo@wip.tu-berlin.de}\\
\hline
\end{tabular}
\caption{Software metadata}
\end{table}

\end{document}